\documentclass[prb,12pt]{revtex4-1}

\usepackage[english]{babel}
\usepackage[intlimits]{amsmath}
\usepackage[dvips]{graphicx}
\usepackage{amsfonts}
\usepackage{amssymb}
\usepackage{graphicx}
\usepackage{bm}

\begin{document}

\begin{titlepage}

\title{Development of Morphogen Gradient: The Role of Dimension and Discreteness}

\author {Hamid Teimouri, Anatoly B. Kolomeisky}
\affiliation{Department of Chemistry and Center for Theoretical Biological Physics, Rice University, Houston, TX 77005-1892}

\begin{abstract}

The fundamental processes of biological development are governed by multiple signaling molecules that create non-uniform concentration profiles known as morphogen gradients. It is widely believed that the establishment of morphogen gradients is a result of complex processes that involve diffusion and degradation of locally produced signaling molecules. We developed a multi-dimensional discrete-state stochastic approach for investigating the corresponding reaction-diffusion models. It provided a full analytical description for stationary profiles and for important dynamic properties such as local accumulation times, variances and mean first-passage times. The role of discreteness in developing of morphogen gradients is analyzed by comparing with available continuum  descriptions. It is found that the  continuum models prediction about  multiple time scales near the source region in two-dimensional and three-dimensional systems is not supported in our analysis. Using  ideas that view the degradation process as an effective potential, the effect of dimensionality on establishment of morphogen gradients is also discussed. In addition, we  investigated how these reaction-diffusion processes are modified with changing the size of the source region.

\end{abstract}

\maketitle

\end{titlepage}

\section{Introduction}

One of the most important biological phenomena is a development of multi-cellular organisms from embryos with initially finite number of genetically identical cells. It is now well established that such spatial patterning and tissue formation is controlled by multiple signaling molecules that are called morphogens.\cite{book,lodish_book,wolpert69,tabata04,crick70}  These signaling molecules create non-uniform concentration profiles, which are also known as morphogen gradients, that serve as local dose-dependent gene regulators for embryo cells. Depending on the concentration of morphogens different genes are turned on or turned off, producing morphologically different cells.\cite{lodish_book,wolpert69} The original ideas of reaction-diffusion control in biological development have been proposed by more than 60 years ago by A. Turing,\cite{turing} and in recent years  a significant number of quantitative studies that uncovered important properties of morphogen gradients in various biological systems have been presented.\cite{porcher10,gregor07,kicheva07,yu09,kerszberg98,entchev00,muller12,drocco11,little11,spirov09,zhou12} However, our understanding of mechanisms of formation of signaling molecules profiles in embryos is still quite limited.\cite{kornberg12} 

To explain complex processes that determine the establishment of morphogen gradients several ideas have been discussed.\cite{kornberg12} But the most popular proposed mechanism is based on a so-called synthesis-diffusion-degradation (SDD) model.\cite{porcher10,drocco12} It assumes that morphogen molecules are synthesized in some localized region of the embryo, then they diffuse along the cellular lines, and with some probability they can associate to receptors on  cells surfaces, from which they are eventually degraded and removed from the system.\cite{kicheva07,yu09,crick70} This model has been widely used, although with a variable success, for analyzing dynamics of signaling molecules in different biological systems.\cite{porcher10,gregor07,kicheva07,yu09,entchev00,drocco11} 

In most cases, theoretical analysis of the morphogen gradient formation employs one-dimensional continuum versions of the SDD model.\cite{berezhkovskii10,berezhkovskii11a,berezhkovskii11b,berezhkovskii11c,berezhkovskii13} A more realistic description of signaling processes that takes into account the structure of embryo requires the application of multi-dimensional models.\cite{mogilner11} The importance of dimensionality has been also pointed out in recent experiments on diffusion of morphogens in extracellular space where geometric obstacles strongly affect trajectories of signaling molecules.\cite{kornberg12,muller13} Recently, the spherically-symmetric continuum SDD model has been investigated for multi-dimensional situations.\cite{gordon13,ellery13} This elegant theoretical method analyzed the kinetics of formation of morphogen gradients, and it also provided analytical expressions for stationary concentration profiles and for local accumulation times (LAT), which are defined as average times to reach locally the steady-state concentrations. One of the most surprising observations of this work is a prediction that for two-dimensional and three-dimensional systems, in contrast to one-dimensional models, there are multiple time scales for dynamics of formation of signaling molecules concentration profiles for the spatial region near the source.\cite{gordon13} It suggested that the dimensionality might play an important role in dynamics of these reaction-diffusion processes.

Analyzing the application of the results from continuum SDD models one has to remember that the continuum picture is an approximation which does not work for all set of parameters. The discrete-state stochastic version of the SDD model on semi-infinite interval has been recently introduced by one of us.\cite{kolomeisky11} The advantage of this approach is that it is valid for {\it all} ranges of the parameter space, and the continuum version is just a special limiting case. It has been also shown that the local accumulation times can be well approximated via the corresponding mean first-passage times, and the degradation process can be viewed as an effective potential that drives morphogens away from the source.\cite{kolomeisky11}

In this paper we develop a discrete-state stochastic framework for description of SDD models with  strongly localized sources in all spatial dimensions. This analysis allows us to compute stationary-state density profiles for morphogens as well as transient dynamic properties such as local accumulations times, mean first-passage times and variance of the local accumulation times. It provides a direct way for measuring the effect of dimensionality in these reaction-diffusion processes. By comparing the obtained results with predictions from continuum models the role of discreteness is also investigated. The existence of multiple times scales in dynamics of morphogen gradients formation for two-dimensional and three-dimensional cases is critically tested, and it is found that there is only one time scale for all distances from the source. In addition, we generalized our approach for the systems with extended source regions, which allowed us to analyze the effect of the producing region size on dynamics of morphogen gradients formation.

The paper is organized as follows. In Sec. II the general $d$-dimensional discrete-state stochastic SDD model is presented and analyzed. The extended version of the model for  source regions of variable size is given  in  Sec. III.  Summary and  concluding remarks are made in the final Sec.IV, while some detailed calculations are presented in Appendices A and B.

\section{A Discrete-State Stochastic SDD Model in General Dimensions}

Let us consider a general discrete-state stochastic SDD model in $d$ spatial dimensions as illustrated in Fig. 1 for $d=2$. Any lattice site in the $d$-dimensional space is characterized by $d$ coordinates, namely $\vec{\bf{n}}=(n_{1},n_{2},...,n_{d})$. We assume that morphogens are produced at the origin  $\vec{\bf{n}}_{0}=(0,0,...,0)$ with a time-independent rate $Q$. Then, from any lattice site $\vec{\bf{n}}=(n_{1},n_{2},...,n_{d})$ they  can jump to any  nearest neighbor site with a diffusion rate $u$. The particle can also be degraded at any position with a rate $k$: see Fig. 1. The continuum limit is realized when $u \gg k$, i.e., when the diffusion rate is much larger than the degradation rate. For convenience, we adopt here a single-molecule view of the process, in which the concentration of signaling molecules is equivalent to the probability of finding a single morphogen particle at given site.\cite{kolomeisky11} We define a function $P(n_{1},n_{2},...,n_{d};t)$ as the probability to find the particle at the position $\vec{\bf{n}}=(n_{1},n_{2},...,n_{d})$ at time $t$.  The temporal evolution of these probabilities is governed by a set of  master equations,
\begin{equation}\label{master_eq1}
 \frac{d P(n_{1},n_{2},...,n_{d};t)}{dt}=u \sum_{nn}{ P(n_{1},n_{2},...,n_{d};t)} - (2ud + k )P(n_{1},n_{2},...,n_{d};t),
\end{equation}
where $\sum_{nn}{}$ is an operator corresponding to summing over all nearest neighbors, namely:
\begin{eqnarray}\label{master_eq2}
\sum_{nn}{ P(n_{1},n_{2},...,n_{d};t)} = P(n_{1}-1,n_{2},...,n_{d};t) + P(n_{1}+1,n_{2},...,n_{d};t) \nonumber \\
+P(n_{1},n_{2}-1,...,n_{d};t) + P(n_{1},n_{2}+1,...,n_{d};t)+....
\end{eqnarray}
 For the origin site we have a slightly different master equation,
\begin{equation}\label{master_beq3}
 \frac{d P(0,0,...;t)}{dt}=Q+u \sum_{nn}{ P(0,0,...;t)}  - (2du+ k)P(0,0,...;t) 
\end{equation}
with
\begin{eqnarray}\label{master_nn1}
\sum_{nn}{ P(0,0,...;t)} = P(-1,0,...,0;t) + P(1,0,...,0;t) \nonumber \\
+P(0,-1,...,0;t) + P(0,1,...,0;t)+...
\end{eqnarray}
At large times, these equations can be solved exactly, producing the stationary density profiles,
\begin{equation}\label{station_stat1}
P^{(s)}(n_{1},n_{2},...,n_{d})=  \frac{2 Q x^{\mid n_{1} \mid + \mid n_{2} \mid +...+ \mid n_{d} \mid }}{ \sqrt{k^{2} + 4duk}}= \frac{2 Q }{ \sqrt{k^{2} + 4duk}}\exp{(\frac{-\mid n_{1} \mid - \mid n_{2} \mid -...- \mid n_{d} \mid }{\lambda})},
\end{equation}
where  
\begin{equation}\label{lambda}
x=(2du+k - \sqrt{k^{2} + 4duk})/(2du), \quad  \lambda=-1/\ln{x},
\end{equation}
and $\lambda$ is a decay length. For $d=1$ these expressions reduce, as expected, to already known results.\cite{kolomeisky11} 

One can see that that in the steady-state the density profile  is the exponentially decaying function with the decay length  being independent of the source production rate $Q$. Similar behavior has been observed in multi-dimensional continuum SDD models.\cite{gordon13,ellery13} In our approach, the continuum limit corresponds to the case when the diffusion rate is much larger than the degradation rate, $u \gg k$. In this case we have $\lambda \simeq \sqrt{(du/k)}$. At another limit, for fast  degradation rates, $k \gg u$, the decay length is equal to $\lambda \simeq 1/\ln{(k/2du)}$. The analysis of Eqs. (\ref{station_stat1}) and (\ref{lambda}) suggests that increasing $d$ leads to lower probability to find the signaling molecules at the origin, while at the same time the decay in the density profile is also slower. There is an important difference between the predictions for the decay length in the continuum and discrete SDD models. We argue that $\lambda$ is generally larger ($\sim \sqrt{d}$ in the continuum limit), and it might be important for interpretation of experimental results in the formation of morphogen gradients.\cite{mogilner11}

\subsection{Local Accumulation Times}

An important dynamic property of morphogen gradients formation are local accumulation times. They are defined as average times at which the stationary density profile is achieved at given spatial position. Berezhkovskii and coworkers\cite{berezhkovskii10} have introduced  a method of calculating explicitly these quantities by using  local relaxation functions $R(n_{1},n_{2},...,n_{d};t)$, which can be written as \begin{eqnarray}\label{relax}
 R(n_{1},n_{2},...,n_{d};t)=\frac{ P(n_{1},n_{2},...,n_{d};t)-P^{(s)}(n_{1},n_{2},...,n_{d})}{ P(n_{1},n_{2},...,n_{d};t=0)-P^{(s)}(n_{1},n_{2},...,n_{d})} \nonumber \\
 =1- \frac{ P(n_{1},n_{2},...,n_{d};t)}{P^{(s)}(n_{1},n_{2},...,n_{d})}
\end{eqnarray}
for the discrete-state multi-dimensional SDD models. The physical meaning of the local relaxation function is that it gives a measure of how close the system to the steady-state conditions. It ranges from $R=1$ at $t=0$ to $R=0$ when the system reaches the stationary state at given location. Introducing the Laplace transform of this function, $\widetilde{ R}(n_{1},n_{2},...,n_{d};s)=\int_{0}^{\infty} {R(n_{1},n_{2},...,n_{d};t) e^{-st}} dt$, it can be shown that the LAT are given by\cite{berezhkovskii10}
\begin{eqnarray}\label{accumulationtime}
 t(n_{1},n_{2},...,n_{d})=-\int_{0}^{\infty} {t \frac{\partial R(n_{1},n_{2},...,n_{d};t)}{ \partial t}} dt} =\int_{0}^{\infty}{R(n_{1},n_{2},...,n_{d};t)dt \nonumber \\
=\widetilde{ R}(n_{1},n_{2},...,n_{d};s=0).
\end{eqnarray}
From this relation the explicit expressions for the local accumulation times can be found:
\begin{equation}\label{accumulationtime2}
 t(n_{1},n_{2},...,n_{d})= \frac{(2du + k)}{(k^{2}+4duk)}  
+ \frac{\mid n_{1} \mid+\mid n_{2} \mid+...+\mid n_{d} \mid}{\sqrt{k^{2}+4duk}}. 
\end{equation}

To compare our results with continuum SDD models (which were analyzed for spherically symmetric conditions),\cite{gordon13,ellery13} it is convenient to consider a specific direction in space along a radial vector $\vec{\bf{r}}=(n_{1},n_{2},...,n_{d})$  where $\mid n_{1} \mid=\mid n_{2} \mid=...=\mid n_{d} \mid$. One can easily show that $\mid n_{1} \mid=\frac{r}{\sqrt{d}}$ where $r$ is the radius of hypersphere enclosing the hypercube of volume $(2\mid n_{1} \mid)^{d}$. This corresponds to a line of length $2\mid n_{1} \mid$, a square of area $4\mid n_{1} \mid^{2}$ and a cube of volume $8\mid n_{1} \mid^{3}$ in one, two and three dimensions respectively. Therefore, the equivalent expression for the LAT at the distance $r$ from the origin is equal to 
\begin{equation}\label{lat_spherical}
 t(r)= \frac{(2du + k)}{(k^{2}+4duk)}  + (\frac{\sqrt{d}}{\sqrt{k^{2}+4duk}})r. 
\end{equation}
In the fast degradation limit, $k \gg u$, this equation simplifies into
\begin{equation}\label{LAT1}
 t(r) \simeq \frac{1}{k}  + \frac{r \sqrt{d}}{k}. 
\end{equation}
In the fast diffusion case, $u \gg k$, we obtain
\begin{equation}\label{LAT2}
 t(r) \simeq \frac{1}{2k}  + \frac{r }{2\sqrt{uk}}. 
\end{equation}

The dependence of the LAT on the radial distance $r$ for 1D, 2D and 3D systems for various sets of parameters is illustrated in Figs. 2-4. In one dimension, the expression for the local accumulation time derived in the discrete-state SDD model reads as
\begin{equation}\label{lat@1d_dsdd}
 t(r) \simeq \frac{2u+k}{k^{2}+4uk} + \frac{r}{\sqrt{k^{2}+4uk}},  
\end{equation}
while in the continuum SDD model it was shown that\cite{gordon13,ellery13}
\begin{equation}\label{lat@1d_csdd}
 t(r)\simeq \frac{1}{2k} + \frac{r}{2\sqrt{uk}}. 
\end{equation}
The last expression could also be obtained in the limit of very large diffusion, $u \gg k$, from Eq. (\ref{lat@1d_dsdd}). These results are plotted in Fig. 2. For fast diffusion rates the predictions from discrete and continuum calculations, as expected, fully agree (see Fig. 2c). The deviations between discrete and continuum models start to appear for comparable diffusion and degradation rates (Fig. 2b), and for  fast degradation rates (Fig. 2a) the local accumulation times for discrete case is smaller for all range of distances except very close to the origin.  In this regime the continuum model cannot be applied, but the discrete-state approach is valid for analyzing  reaction-diffusion processes of morphogen gradients formation.

Similar calculations in two dimensions yield the following expression for the LAT,
\begin{equation}\label{lat@2d_dsdd}
 t(r) \simeq \frac{4u+k}{k^{2}+8uk} + \frac{\sqrt{2}r}{\sqrt{k^{2}+8uk}}.
\end{equation}
The 2D continuum SDD model predicts the following result,\cite{ellery13}
\begin{equation}\label{lat@2d_csdd}
 t(r)\simeq \frac{r}{2\sqrt{uk}} \ \frac{K_{1}(r\sqrt{k/u})}{K_{0}(r\sqrt{k/u})}, 
\end{equation}
where $K_{m}(x)$ is the $m$-th order modified Bessel function of the second kind. Note that taking the limit of $u \gg k$ in our theoretical approach in  Eq. (\ref{lat@2d_dsdd}), which supposed to be corresponding to the continuum limit, produces a different expression,
\begin{equation}
 t(r) \simeq \frac{1}{2k} + \frac{r}{2\sqrt{uk}}.
\end{equation}                    
Fig. 3 presents these functions for different sets of parameters. We can see that even for large diffusion rates (Fig. 3c) the predictions of discrete and continuum models do not fully agree, but for large distances from the source the differences are small. Again, as for 1D case, the deviations between two approaches start to build up with decreasing the diffusion rate (Fig. 3b), and  for large degradation rates the LAT for the discrete-state model are significantly smaller for most distances, except for very small $r$ (Fig. 3a).

For 3D systems the expressions for the local accumulation times in the discrete SDD model is given by
\begin{equation}\label{lat@3d_dsdd}
 t(r) \simeq \frac{6u+k}{k^{2}+12uk} + \frac{\sqrt{3}r}{\sqrt{k^{2}+12uk}}.
\end{equation}
The continuum description of the same reaction-diffusion processes yields,\cite{gordon13,ellery13}
\begin{equation}\label{lat@3d_csdd}
 t(r)\simeq \frac{r}{2\sqrt{uk}}. 
\end{equation}
For this case, the LAT are presented in Fig. 4. The observed trends are very similar to 2D systems, but with stronger deviations between discrete and continuum predictions. Again, even in the continuum limit our theoretical predictions for the LAT do not agree with calculations from continuum SDD models,\cite{gordon13,ellery13} although for large $r$ the differences are not significant. 

Comparing local accumulation times for discrete-state and continuum SDD models, the important observation can be made  that for all regimes the continuum models in both 2D and 3D predict  $t(r=0)=0$, while in the discrete-state analysis this time is always finite. Since at $t=0$ there are no morphogens in the system and the LAT is associated with the time to reach the stationary density at given position, it is expected that this quantity to be finite even at the origin. It seems that predictions of the continuum models do not satisfy this requirement for $d>1$, suggesting that they cannot properly describe reaction-diffusion processes of formation of signaling molecules profiles close to the origin, even for conditions when the continuum approximation should hold. No such problems exist for the discrete-state approach. This is the main reason for predicting multiple time scales in the continuum description  (for $d>1$) of the development of signaling molecules profiles. In the discrete model there is one time scale, given by the LAT, at all distances. The possible reason for this discrepancy might be special boundary conditions utilized in solving differential equations that describe these reaction-diffusion processes in the continuum approach.\cite{gordon13,ellery13}

It is interesting also to investigate the role of dimensionality in the establishment of morphogen gradients. For fast degradation the discrete model predictions are given by Eq. (\ref{LAT1}), while in the fast diffusion limit they are given by Eq. (\ref{LAT2}). The dependencies of the local accumulation times on dimension $d$ for the discrete and continuum  SDD models are plotted in Fig. 5 for the  sites that are far away from the source ($r \gg 0$), and in Fig. 6 for the sites that are close to the origin ($r=0$). Surprisingly, the results are quite different. For fast degradation rates, the LAT is increasing with $d$ for the sites not so close to the source, while at the origin and closest sites the LAT is slowly decreasing (compare upper plots in Figs. 5 and 6). A similar behavior is observed  for comparable diffusion and degradation rates, although the effect is getting weaker (see middle plots in Figs. 5 and 6). For continuum limit, $u \gg k$, the LAT in both positions become independent of the dimension, as correctly predicted by Eq. (\ref{LAT2}). 

The following arguments can be given to understand this behavior in the discrete SDD model. At $t=0$ the signaling molecules start at the origin, $r=0$. The local accumulation time is the average time to reach the steady-state density at a given position, so it depends on possible pathways connecting the origin and any site at $r>0$. Increasing the dimensionality produces more pathways so it takes  longer time if the diffusion is the rate limiting step. For this reason, the LAT depends on $d$  for diffusion rates comparable or smaller the degradation, while for $u \gg k$ there is no dependence on the dimension - the degradation is a rate-limiting step in this case. At sites close to the origin these diffusion pathways do not play any role. But the stationary density at these sites  is also smaller for larger $d$, so it is faster to reach the steady-state concentration with increasing $d$ when the diffusion is rate limiting.

\subsection{Mean First-Passage Times}

It has been argued before that in order to understand mechanisms of formation of morphogen gradients it is useful to consider mean first-passage times (MFPT) to reach specific locations for molecules starting from the origin.\cite{kolomeisky11} The reason for this is the fact that first-passage events are the dominating factors determining the local accumulation times at large distances, at least for one-dimensional systems,\cite{kolomeisky11}; the explicit connections between these quantities have been recently studied  for $d=1$.\cite{berezhkovskii11b} It is important to understand if first-passage processes describe the morphogen gradient formation  in higher dimensions.

To compute MFPT we define $f(n_{1},n_{2},...,n_{d};t)$ to be a first-passage probability to reach for the first time the site $\vec{\bf{n}}=(n_{1},n_{2},...,n_{d})$ at time $t$ if at $t=0$ the particle started at the origin. The temporal evolution of this function follows a backward master equation,\cite{redner}
\begin{equation}\label{master_mfpt1}
 \frac{d f(n_{1},n_{2},...,n_{d};t)}{dt}=u \sum_{nn}{ f(n_{1},n_{2},...,n_{d};t)} - (2ud + k )f(n_{1},n_{2},...,n_{d};t),
\end{equation}
where $\sum_{nn}{}$ is the operator that sums over all nearest neighbors. Utilizing the Laplace transformations, we obtain
\begin{equation}\label{mfpt0}
\widetilde{f}(n_{1},n_{2},...,n_{d};s)= \frac{ 2 \sqrt{a^{2}-4d^{2}u^{2}} y^{\mid n_{1} \mid + \mid n_{2} \mid + ...+ \mid n_{d} \mid }} {(a-2du+\sqrt{a^{2}-4d^{2}u^{2}})y^{2 \mid n_{1} \mid + 2 \mid n_{2} \mid + ... + 2 \mid n_{d} \mid} - (a-2du-\sqrt{a^{2}-4d^{2}u^{2}})}.
\end{equation}
where 
\begin{equation}
a=s+2du+k, \quad y=\left[a+\sqrt{a^{2}-4d^{2}u^{2}}\right]/2ud.
\end{equation}
The conditional mean  first-passage time to reach the site $\vec{\bf{n}}=(n_{1},n_{2},...,n_{d})$ can be found from the following expression,
\begin{equation}
\tau(n_{1},n_{2},...,n_{d})=-\frac{\frac{d\widetilde{f}(n_{1},n_{2},...,n_{d};s)}{ds}_{|_{s=0}}}{\widetilde{f}(n_{1},n_{2},...,n_{d};s)_{|_{s=0}}}
\end{equation}
After some algebra, the corresponding expression for the MFPT is derived,
\begin{eqnarray}\label{mfpt_single}
&& \tau(n_{1},n_{2},...,n_{d})=\frac{1}{\sqrt{k^{2}+4dku}}[-\frac{2du+k}{\sqrt{k^{2}+4dku}} \nonumber\\
&& + \frac{(2du+k+\sqrt{k^{2}+4dku})z^{\mid n_{1} \mid+\mid n_{2} \mid+...+\mid n_{d} \mid}+(2du+k-\sqrt{k^{2}+4dku})z^{-\mid n_{1} \mid-\mid n_{2} \mid-...-\mid n_{d} \mid}}{(k+\sqrt{k^{2}+4dku})z^{\mid n_{1} \mid+\mid n_{2} \mid+...+\mid n_{d} \mid}-(k-\sqrt{k^{2}+4dku})z^{-\mid n_{1} \mid-\mid n_{2} \mid-...-\mid n_{d} \mid}}\nonumber\\
&& + \frac{(\mid n_{1} \mid+\mid n_{2} \mid+...+\mid n_{d} \mid)(2du+k+\sqrt{k^{2}+4dku})z^{\mid n_{1} \mid+\mid n_{2} \mid+...+\mid n_{d} \mid}}{(k+\sqrt{k^{2}+4dku})z^{\mid n_{1} \mid+\mid n_{2} \mid+...+\mid n_{d} \mid}-(k-\sqrt{k^{2}+4dku})z^{-\mid n_{1} \mid-\mid n_{2} \mid-...-\mid n_{d} \mid}}
\nonumber\\
&& + \frac{(\mid n_{1} \mid+\mid n_{2} \mid+...+\mid n_{d} \mid)(2du+k-\sqrt{k^{2}+4dku})z^{-\mid n_{1} \mid-\mid n_{2} \mid-...-\mid n_{d} \mid}}{(k+\sqrt{k^{2}+4dku})z^{\mid n_{1} \mid+\mid n_{2} \mid+...+\mid n_{d} \mid}-(k-\sqrt{k^{2}+4dku})z^{-\mid n_{1} \mid-\mid n_{2} \mid-...-\mid n_{d} \mid}}].\nonumber\\
\end{eqnarray}
where $z=\left[2du+k+\sqrt{k^{2}+4duk}\right]/2du$. 
For fast degradation rate, $k \gg u$, we obtain a much simpler expression, 
\begin{equation}\label{accumulationtime3}
 \tau(n_{1},n_{2},...,n_{d}) \simeq \frac{\mid n_{1} \mid+\mid n_{2} \mid+...+\mid n_{d} \mid+1}{k},
\end{equation}
which for large radial distances, $r \gg 1$, can also be written as
\begin{equation}\label{MFPT1}
 \tau(r) \simeq \frac{r \sqrt{d}}{k}.
\end{equation}
One can see that this expression agrees well with Eq. (\ref{LAT1}) at large $r$. In the opposite limit of the fast diffusion rates (continuum limit), $u \gg k$, one can show that the MFPT are equal to
\begin{equation}\label{accumulationtime4}
\tau(n_{1},n_{2},...,n_{d})\simeq \frac{\mid n_{1} \mid+\mid n_{2} \mid+...+\mid n_{d} \mid+1}{2\sqrt{kud}}.
\end{equation}\label{MFPT2}
For $r \gg 1$ it modifies into
\begin{equation}
\tau(r)\simeq \frac{r}{2\sqrt{uk}},
\end{equation}
which asymptotically agree with Eq. (\ref{LAT2}) at large distances. These results again support the idea that main contribution to the LAT at large distances from the origin are due to the MFPT, extending the validity of this idea to all dimensions. This is an important observation since the first-passage analysis is a well developed mathematical tool that was already successfully employed for analyzing multiple physical, chemical and biological processes.\cite{redner}

To support arguments about the importance of the first-passage events in dynamics of the morphogen gradient development, the ratio of MFPT over LAT is plotted in Fig. 7 for different sets of parameters. One can see that this ratio is always approaching 1 for large distances. Larger degradation rates as well as higher dimensions lead to faster converging to unity, while in the continuum limit (fast diffusion rates) the effect of dimension disappears.

\subsection{Effective Potentials}

Analyzing mechanisms of morphogen gradient formation suggested a new idea that degradation can be viewed as an effective potential that drives the signaling molecules away from the source.\cite{kolomeisky11} Thus morphogens are not simply diffusing with equal probability in each direction, but their motion is biased by this effective potential to move further away from the source. This concept can be extended and applied for the multi-dimensional SDD models of creating signaling molecules profiles.

The effective potential can be easily calculated from the  stationary profile, leading to
\begin{eqnarray}\label{effp1}
U_{eff}(n_{1},n_{2},...,n_{d})\simeq k_{B}T \ln{P^{(s)}(n_{1},n_{2},...,n_{d})} \nonumber\\
= k_{B}T(\mid n_{1} \mid + \mid n_{2} \mid +...+ \mid n_{d} \mid ) \ln x,
\end{eqnarray}
and it can be rewritten as follows (for $i=1,2,...,d$),
\begin{equation}\label{effp2}
U_{eff}(n_{1},n_{2},...,n_{d}) \simeq  \sum_{i=1}^{d} U_{eff}(n_{i}), \quad U_{eff}(n_{i})=k_{B}T \mid n_{i} \mid \ln x.
\end{equation}
This equation has an important physical meaning suggesting that the overall potential is a sum of potentials along each of the coordinate axes. Consequently, in higher dimensions the effective potential is stronger than one dimension. The reason for this is that in higher dimensions morphogens can diffuse in more directions and thus the probability of returning to the origin decreases as $\sim 1/d$. It also suggests that there is a constant force component, 
\begin{equation}
F_{i}=-\frac{\partial U_{eff}}{\partial \mid n_{i} \mid}=- k_{B}T \ln x= k_{B}T/\lambda,
\end{equation}
along  each axis that drives signaling molecules away from the source.

The importance of this concept can be seen in explaining most dynamic properties of morphogen reaction-diffusion systems. The linear dependence of the LAT on distances from the sources [see Eq. (\ref{accumulationtime2})] is the consequence of the effective potential that changes the unbiased diffusion of morphogen molecules into a driven motion. Similarly, the linear dependencies of the MFPT on distances have the origin: see Eq.(\ref{mfpt_single}). It also provides an alternative explanation for dependence of the LAT on dimension for sites near the source (Fig. 6): increasing $d$ make this potential stronger so it drives particles faster to their destinations. The same reasoning can be used to understand why the MFPT approximate the LAT better at higher dimensions or at faster degradations (Fig. 7).

\subsection{Variance of Local Accumulation Times}

 The advantage of presented theoretical method is that it allows us to calculate all dynamic properties of the morphogen gradient formation. To illustrate this, let consider higher moments of the local accumulation times. The LAT itself is the first moment as indicated in Eq. (\ref{accumulationtime}). The second moment, which is a mean-squared local accumulation time, can be also calculated from the local relaxation function,
\begin{equation}\label{variance2}
<t^{2}(n_{1},n_{2},...,n_{d})>= - \int_{0}^{\infty}t^{2} \frac{d R(n_{1},n_{2},...,n_{d};t)}{dt}dt=-2{\frac{d\widetilde{R}(n_{1},n_{2},...,n_{d};s)}{ds}}_{|s=0}.
\end{equation}
Substituting the explicit expression for $\widetilde{R}(n_{1},n_{2},...,n_{d};s)$ we obtain,
\begin{eqnarray}\label{variance3}
 <t^{2}(n_{1},n_{2},...,n_{d})>= \frac{2(\mid n_{1} \mid+\mid n_{2} \mid+...+\mid n_{d} \mid)^{2}-2} {(k^{2}+4duk)}  \nonumber \\
+ \frac{4(2du + k)(\mid n_{1} \mid+\mid n_{2} \mid+...+\mid n_{d} \mid)}{(k^{2}+4duk)^{3/2}} +\frac{6(2du + k)^{2}}{(k^{2}+4duk)^{2}}. 
\end{eqnarray}
It can be shown that generally  the $m$-th moment of the LAT is given by
\begin{equation}\label{variance1}
<t^{m}(n_{1},n_{2},...,n_{d})>=(-1)^{m-1}m{\frac{d^{m-1}\widetilde{R}(n_{1},n_{2},...,n_{d};s)}{ds^{m-1}}}_{|s=0}.
\end{equation}
The explicit forms for the first and second moments of the LAT allow us to calculate a variance, which gives a measure of fluctuations in the local accumulation times. The variance of the local accumulation time is equal to
\begin{eqnarray}
\sigma(n_{1},n_{2},...,n_{d})=\sqrt{<t^{2}>-<t>^{2}}= \left[ \frac{(\mid n_{1} \mid+\mid n_{2} \mid+...+\mid n_{d} \mid)^{2}-2}{(k^{2}+4duk)}    \right.      \nonumber \\
\left. +\frac{2(2du + k)(\mid n_{1} \mid+\mid n_{2} \mid+...+\mid n_{d} \mid)}{(k^{2}+4duk)^{3/2}}+ \frac{5(2du + k)^{2}}{(k^{2}+4duk)^{2} } \right]^{1/2}.
\end{eqnarray}
In terms of the radial distance $r$ the variance can be written as
\begin{equation}
\sigma(r)=\left[ \frac{d r^{2}-2}{(k^{2}+4duk)}+ \frac{2 r \sqrt{d} (2du + k)}{(k^{2}+4duk)^{3/2}}+ \frac{5(2du + k)^{2}}{(k^{2}+4duk)^{2} } \right]^{1/2}.
\end{equation}
In the limit of fast diffusion rates (continuum  limit) the expression for the variance is simpler,
\begin{equation}
\sigma(r) \simeq \frac{\sqrt{5}}{2k}+\frac{r}{2\sqrt{5uk}}.
\end{equation}
This result implies that the variance becomes independent of the dimension  for $u \gg k$. For the case of the fast degradation rates ($k \gg u$) the variance behavior is different,
\begin{equation}
\sigma(r) \simeq \frac{\sqrt{d r^{2}+2 r \sqrt{d}+3}}{k}.
\end{equation} 
In this limit the variance increases with $d$ but becomes independent of the diffusion rate.

The variances normalized with respect to the LAT are presented in Fig. 8. We can see that at large distance the ration $\sigma(r)/t(r)$ is always approaching unity. The increase in the degradation rates lowers the variance, while increasing the diffusion rate  make the system more noisy. At fast degradation rates, increasing the dimension lowers the variance (Fig. 8a), while for large diffusion rates there is no dependence on $d$. The importance of these observations is that they suggest possible ways of how nature might control noise in morphogen gradient systems.

\section{A Discrete-State Stochastic SDD Model with Extended Source Region}

In the model discussed before it was assumed that the source of signaling molecules is sharply localized at the origin. In real systems the morphogen production is more delocalized,\cite{book,lodish_book,porcher10} and it raises a question of how the size of the source region affects the dynamics of morphogen gradient formation. To answer this question, the original $d$-dimensional  discrete-state stochastic SDD model is generalized to take into account this effect by assuming that morphogens can be produced from the discrete point sources distributed inside a hypercubic of volume $(2R)^{d}$ in $d$-dimensional space. This corresponds to a line of length $2R$, a square of area $4R^{2}$ and a cube of volume $8R^{3}$ in one, two and three dimensions respectively. It is assumed that the production rate at each site is equal to $Q$.

It is convenient to introduce a propagator function  $G(m_{1},m_{2},...,m_{d};t_{0}|n_{1},n_{2},...,n_{d};t)$ defined as the conditional probability to find the particle at site $\vec{\bf{n}}=(n_{1},n_{2},...,n_{d})$  if it starts at $t_{0}$ at site $\vec{\bf{m}}=(m_{1},m_{2},...,m_{d})$.\cite{berezhkovskii11a}  Then the probability $P(n_{1},n_{2},...,n_{d};t)$ of finding the particle at site $\vec{\bf{n}}=(n_{1},n_{2},...,n_{d})$ at time $t$ can be expressed as a superposition of the corresponding propagators:
\begin{equation}\label{g1}
P(n_{1},n_{2},...,n_{d};t)=\sum_{m_{1}=-R}^{R}\sum_{m_{2}=-R}^{R}...\sum_{m_{d}=-R}^{R}{\int_{0}^{t}
{ G(m_{1},m_{2},...,m_{d};t_{0}|n_{1},n_{2},...,n_{d};t)}} dt_{0} ,
\end{equation}
where the summation is performed over the region of space where particle sources are located. The corresponding master equations for temporal evolution of probabilities can be solved in the large-time limit (see Appendix A), leading to
\begin{equation}\label{station_conti}
P^{(s)}(n_{1},n_{2},...,n_{d})=   \frac{2 Q\Gamma^{d} }{\sqrt{k^{2}+4duk}}\exp{(\frac{-\mid n_{1} \mid - \mid n_{2} \mid -...- \mid n_{d} \mid}{\lambda} +\frac{dR}{\lambda})},
\end{equation}
with $x$ and $\lambda$ defined in Eq. (\ref{lambda}), while a new function $\Gamma$ is given by
\begin{equation}
\Gamma=\frac{x^{R+1}+x^{R}-2}{{x-1}}.
\end{equation}
It characterizes  the extended source region. In the case of $R=0$ we get $\Gamma=1$ and the results of Sec. II are fully recovered. In general, $\Gamma$ can be a complex function that strongly depends on geometry and distribution of morphogen sources. 

To simplify calculations, here we assumed that the morphogen molecules are produced in the hypercubic region around the origin, but our analysis can be easily extended to geometrically more complex source regions. It is also important to note  that our model differ from the continuum SDD models where morphogens can be produced only at the surface of the region of size $R$.\cite{gordon13,ellery13} In our case, which is much closer to the real situation, the sources of the signaling molecules are distributed inside of the production region, and morphogens can a diffuse in the source area, cross  into non-productive region and return back. 

The local accumulation times for the discrete SDD model with the extended source can be calculated following the same procedure as explained in Sec. II. It yields
\begin{eqnarray}
 t(n_{1},n_{2},...,n_{d})= \frac{1}{\sqrt{k^{2}+4duk}}\left[\frac{2du+k}{\sqrt{k^{2}+4duk}} + \mid n_{1} \mid+\mid n_{2} \mid+...+\mid n_{d} \mid- dR \right. & \nonumber \\
\left. +d\frac{(R+1)x^{R+1}+Rx^{R}-x\Gamma}{\Gamma(x-1)} \right], &
\end{eqnarray}
which at the distance $r$ (along the vector $\vec{\bf{r}}=(n_{1},n_{2},...,n_{d})$  with $\mid n_{1} \mid=\mid n_{2} \mid=...=\mid n_{d} \mid$) is modified into
 \begin{eqnarray}\label{lat_rsys}
 t(r)= \frac{d}{\sqrt{k^{2}+4duk}}\left[\frac{2du+k}{d\sqrt{k^{2}+4duk}} +\frac{r}{\sqrt{d}}-R +\frac{(R+1)x^{R+1}+Rx^{R}-x\Gamma}{\Gamma(x-1)}\right].
\end{eqnarray}
In the limit of fast degradation rates, $k \gg u$, the resulting expression is simpler,
\begin{equation}\label{LAT2_multiple}
 t(r) \simeq \frac{1}{k}  + \frac{d}{k}(\frac{r}{\sqrt{d}}-R), 
\end{equation}
which for $R=0$ reduces, as expected, to Eq. (\ref{LAT1}). For the special position on the surface of the producing area at the edge of the hypercube, $r=R\sqrt{d}$, it predicts even simpler expression $t(R\sqrt{d}) \simeq 1/k$. For fast diffusion rates ($u \gg k$) one can show that the LAT is given by
\begin{equation}
t(r) \simeq \frac{1}{2k} + \frac{\sqrt{d}}{2\sqrt{uk}} \left[ (\frac{r}{\sqrt{d}}-R)+\frac{2R^{2}}{2R+1} \right],
\end{equation}
which for $R=0$ reduces to Eq. (\ref{LAT2}).

The LAT at the surface of the production area for different dimensions are plotted in Fig. 9. One can see that for large degradation rates the LAT become independent of $d$ (Fig. 9a), while for larger diffusion rates there is a dependence on the dimensionality. It can be explained using the following arguments. At very large $k$, fluxes from other source sites do not reach this specific location - particles are degraded before they can diffuse to neighboring sites. In this case the LAT should not depend on $R$ and on $d$ - only local dynamics at the given site is important. For faster diffusion (Fig. 9b and 9c) the role of fluxes from neighboring sites becomes more important so the dependence on $R$ will show up. But the contribution form the sites that are further away will be much smaller. As a result there is a saturation behavior at $R \gg 1$. In addition, increasing the dimension leads to higher stationary concentrations so it takes more time to reach the steady-state conditions, and this explains why the LAT are the highest for 3D and the lowest for 1D systems.

Our analysis can also be extended for computation of mean first-passage times for the systems with extended production volume. The explicit formulas for MFPT are quite bulky and they are fully derived in Appendix B. Here we present simpler limiting expressions. For slow degradation (conditions close to the continuum limit)  we obtain 
\begin{equation}
\tau(n_{1},n_{2},...,n_{d})\simeq(\mid n_{1} \mid+\mid n_{2} \mid+...+\mid n_{d} \mid  +1 - dR)/2\sqrt{kud},
\end{equation}
which for the single-point source ($R=0$) reduces to Eq. (\ref{accumulationtime4}). For the opposite limit of slow diffusion rates, $k \gg u$, the MFPT depends only on the degradation rate,
\begin{equation}
\tau(n_{1},n_{2},...,n_{d}) \simeq (\mid n_{1} \mid+\mid n_{2} \mid+...+\mid n_{d} \mid +1 - dR)/k, 
\end{equation}
which also for the case of $R=0$ is identical to Eq. (\ref{accumulationtime3}). In both limiting cases MFPT decrease for larger production areas since there are source sites closer to the given position.

Varying the size of the production region modifies also the effective potential that morphogen molecules experience in the system due to degradation. From the stationary densities we obtain, 
\begin{equation}
U_{eff}(n_{1},n_{2},...,n_{d})\simeq k_{B}T\Gamma^{d}(\mid n_{1} \mid + \mid n_{2} \mid +...+ \mid n_{d} \mid - dR) \ln x,
\end{equation}
which can be rewritten as
\begin{equation}
U_{eff}(n_{1},n_{2},...,n_{d}) \simeq  \sum_{i=1}^{d} \left[ U_{eff}(n_{i})-U_{eff}(R)\right],
\end{equation}
where we defined (for $i=1,,2...,d$)
\begin{equation}
U_{eff}(n_{i})= k_{B}T \Gamma^{d} \mid n_{i} \mid \ln x, \quad U_{eff}(R)=k_{B}T \Gamma^{d} R \ln x.
\end{equation}
The corresponding force along the axis $i$ that effectively pushes morphogens away from the source  is given by
\begin{equation}
F_{i}=-\frac{\partial U_{eff}}{\partial \mid n_{i} \mid}=- k_{B}T \Gamma^{d} \ln x= k_{B}T \Gamma^{d}/\lambda,
\end{equation}
leading to stronger forces with increasing the size of the producing area.

\section{Summary and Concluding Remarks}

We developed a multi-dimensional discrete-state stochastic theoretical framework for understanding reaction-diffusion processes of morphogen gradients formation. The approach provides a full analytical description of stationary state and dynamic properties of complex systems where signaling molecules profiles are created. It allowed us to fully analyze the role of discreteness by comparing with current continuum theoretical models, as well as the effect of the dimensionality. 

It is found that at large times the system will reach stationary exponential density profiles with the decay length that increases with the dimension, in contrast to the continuum methods which predict the decay length to be independent of $d$.  The differences between two approaches become larger in analyzing dynamic properties such as the local accumulation times that describe the relaxation to the stationary-state behavior. Continuum models predict that the LAT is approaching zero at the source, resulting in multiple time scales that control dynamics of the system. In contrast, our calculations suggest that the local accumulation times are always finite and they provide the only time scale to describe the kinetics of morphogen gradients formation. Thus, it is argued that current continuum models cannot be used in analyzing these complex reaction-diffusion dynamics  at distances closer to the source, while our discrete approach does not have any problems.

From the presented discrete method an interesting dependence of dynamic properties on dimensions is observed. It is found that for sites close to the source, when the degradation is faster than the diffusion, the LAT times decrease with the dimension, while for regions far away form the source the dependence is reversed. At the same time, for large diffusion rates no effect is observed at any distance. It is explained by accounting for possible pathways connecting the source and the given location in the system. We also analyzed another dynamic property, mean first-passage times.  It is shown that at large distances from the source the MFPT provide an excellent approximation for the LAT, and the approximation is better for higher dimensions and larger degradation rates, while at the continuum limit (fast diffusion) there is no dependence on $d$ and the approximation works not as well.

The concept that degradation processes can be viewed as an effective potential that pushes signaling molecules away from the source has been extended to multi-dimensional systems. It is found that increasing the dimension makes this potential stronger, and this simple idea was powerful enough to explain most trends in dynamic properties, such as the linear dependence of the LAT and MFPT on the distances and the effect of dimensions. In addition, the method allowed us to compute higher moments of the local accumulation times, and specific calculations have been made for estimation of variances. Finally, we extended our method for analyzing reaction-diffusion systems with variable range of production region by explicitly estimating all dynamic and stationary properties and their dependencies of the size of the source volume and dimensions. 

It was argued that the presented discrete-state stochastic approach allows to capture all relevant physical-chemical properties of the development of morphogen gradients. The  main success of the method is a full analytical description of all involved processes at all times and distances. Another advantage is that other biochemical and biophysical processes can be consistently incorporated.  For example, it will be crucial to extend the models to take into account more complex phenomena such as non-uniform production rates,  cooperative mechanisms of degradation and possible bindings/unbindings  of morphogens to other molecules in the system.  It will be also very important to test these theoretical ideas in experimental studies.

\section*{Acknowledgments}

We would like to acknowledge  the support from the Welch Foundation (grant C-1559).

\section*{Appendix A: Calculations of Stationary Density Profiles for the System with Extended  Production Volume}

In Sec. III we introduced the propagator or Green function, $G(m_{1},m_{2},...,m_{d};t_{0}=0|n_{1},n_{2},...,n_{d};t)$, that  characterizes  the system with extended range of morphogen production. Its temporal evolution is governed by the following master equation:
\begin{eqnarray}\label{green1}
 \frac{dG(m_{1},m_{2},...,m_{d};t_{0}|n_{1},n_{2},...,n_{d};t)}{dt}-u \sum_{nn}{ G(m_{1},m_{2},...,m_{d};t_{0}|n_{1},n_{2},...,n_{d};t)} \nonumber \\
 + (2ud + k )G(m_{1},m_{2},...,m_{d};t_{0}|n_{1},n_{2},...,n_{d};t)=Q \delta(n_{1}-m_{1}) \delta(n_{2}-m_{2})...\delta(n_{d}-m_{d}),
  \nonumber \\
\end{eqnarray}
where $\delta(x)$ is a Kronecker delta and $\sum_{nn}{}$ is the operator that sums over the nearest neighbors,
\begin{eqnarray}\label{green2}
\sum_{nn}{ G(m_{1},m_{2},...,m_{d};t_{0}|n_{1},n_{2},...,n_{d};t)} = G(m_{1},m_{2},...,m_{d};t_{0}|n_{1}-1,n_{2},...,n_{d};t) \nonumber \\
+ G(m_{1},m_{2},...,m_{d};t_{0}|n_{1}+1,n_{2},...,n_{d};t) 
+G(m_{1},m_{2},...,m_{d};t_{0}|n_{1},n_{2}-1,...,n_{d};t)+...\nonumber \\
\end{eqnarray}
Eq. (\ref{green1}) can be rewritten as
\begin{equation}\label{operator}
\hat{L}G(m_{1},m_{2},...,m_{d};t_{0}|n_{1},n_{2},...,n_{d};t)=\delta(n_{1}-m_{1})\delta(n_{2}-m_{2})...\delta(n_{d}-m_{d})
\end{equation}
with $\hat{L}$ defined as an operator acting on the Green function. Here, the Green function can be regarded as an auxiliary function which satisfies the appropriate boundary conditions generated by  a singularly point  source located at  $\vec{\bf{m}}=(m_{1},m_{2},...,m_{d})$. The corresponding Green function for the steady state reads then
\begin{equation}\label{green3}
G^{(s)}(m_{1},m_{2},...,m_{d};t_{0}|n_{1},n_{2},...,n_{d};t \rightarrow \infty)=  \frac{2 Q x^{\mid n_{1} \mid -\mid m_{1} \mid +\mid n_{2} \mid- \mid m_{2} \mid +...+ \mid n_{d} \mid- \mid m_{d}\mid }}{\sqrt{k^{2}+4duk}}.
\end{equation}
Now we can calculate probability function  defined in Eq. (\ref{g1}) by summing over the source region,
\begin{equation}\label{green4}
P^{(s)}(n_{1},n_{2},...,n_{d})=\frac{2Q x^{\mid n_{1} \mid +\mid n_{2} \mid +...+ \mid n_{d} \mid}}{\sqrt{k^{2}+4duk}}\sum_{m_{1}=-R}^{R}\sum_{m_{2}=-R}^{R}...\sum_{m_{d}=-R}^{R}{{ x^{ -\mid m_{1} \mid - \mid m_{2} \mid -...-\mid m_{d}\mid }}}.
\end{equation}
The summation over $m_{1},...m_{d}$ can be performed in the following way,
\begin{equation}\label{green5}
\sum_{m_{1}=-R}^{R}x^{-\mid m_{1}\mid}=2\sum_{m_{1}=0}^{R}x^{-m_{1}}-1=\frac{x^{-R}(x^{R}+x^{R+1}-2)}{x-1}.
\end{equation}
Substituting this result into Eq.(\ref{green4}) yields
\begin{equation}\label{green6}
P^{(s)}(n_{1},n_{2},...,n_{d})=\frac{2Q x^{\mid n_{1} \mid +\mid n_{2} \mid +...+ \mid n_{d} \mid- dR }}{\sqrt{k^{2}+4duk}}(\frac{x^{R}+x^{R+1}-2}{x-1})^{d}=\frac{2Q\Gamma^{d} x^{\mid n_{1} \mid +\mid n_{2} \mid +...+ \mid n_{d} \mid-dR  }}{\sqrt{k^{2}+4duk}}.
\end{equation}

\section*{Appendix B: Calculations of Mean First-Passage Times for the System with Extended  Production Volume}

Similarly to the approach explain in Appendix A, we define $F(m_{1},m_{2},...,m_{d};t_{0}|n_{1},n_{2},...,n_{d};t)$ as a  first-passage conditional probability to reach for  the first time the site $\vec{\bf{n}}=(n_{1},n_{2},...,n_{d})$  if  at $t_{0}$ the particle starts at $\vec{\bf{m}}=(m_{1},m_{2},...,m_{d})$.
\begin{eqnarray}\label{green_mfpt1}
 \frac{dF(m_{1},m_{2},...,m_{d};t_{0}|n_{1},n_{2},...,n_{d};t)}{dt}=u \sum_{nn}{ F(m_{1},m_{2},...,m_{d};t_{0}|n_{1},n_{2},...,n_{d};t)} \nonumber \\
 - (2ud + k )F(m_{1},m_{2},...,m_{d};t_{0}|n_{1},n_{2},...,n_{d};t).
  \nonumber \\
\end{eqnarray}
Here again $\sum_{nn}{}$ is the sum operator explained in Eq. (\ref{green2}). The corresponding mean first-passage probability can be expressed as a sum over these propagators,
\begin{equation}\label{green_mfpt2}
f(n_{1},n_{2},...,n_{d};t)=\sum_{m_{1}=-R}^{R}\sum_{m_{2}=-R}^{R}...\sum_{m_{d}=-R}^{R}\int_{0}^{t}{F(m_{1},m_{2},...,m_{d};t_{0}|n_{1},n_{2},...,n_{d};t)dt_{0}}.
\end{equation}
It can be shown that
\begin{eqnarray} \label{green_mfpt3}
\widetilde{f}(n_{1},n_{2},...,n_{d};s)= & &  \nonumber \\
\frac{ 2 \sqrt{a^{2}-4d^{2}u^{2}} y_{1}^{\mid n_{1} \mid + \mid n_{2} \mid + ...+ \mid n_{d} \mid - dR  }} {\Theta_{1}^{d}(a-2du+\sqrt{a^{2}-4d^{2}u^{2}})y_{1}^{2 \mid n_{1} \mid + 2 \mid n_{2} \mid + ... + 2 \mid n_{d} \mid-2d R } - \Theta_{2}^{d}(a-2du-\sqrt{a^{2}-4d^{2}u^{2}})}, & &
\end{eqnarray}
where we defined
\begin{equation}
\Theta_{1}=(\frac{y_{1}^{R+1}+y_{1}^{R}-2}{{y_{1}-1}}), \quad \Theta_{2}=(\frac{y_{2}^{R+1}+y_{2}^{R}-2}{{y_{2}-1}}),
\end{equation}
and
\begin{equation}
y_{1}=\frac{(a+\sqrt{a^{2}-4d^{2}u^{2}})}{2du}, \quad y_{2}=\frac{(a-\sqrt{a^{2}-4d^{2}u^{2}})}{2du}, \quad a=s+2du+k.
\end{equation}

The MFPT to reach the site $\vec{\bf{n}}=(n_{1},n_{2},...,n_{d})$ , can be found from the following expression,
\begin{equation}\label{taun}
\tau(n_{1},n_{2},...,n_{d})=-\frac{\frac{d\widetilde{f}(n_{1},n_{2},...,n_{d};s)}{ds}_{|_{s=0}}}{\widetilde{f}(n_{1},n_{2},...,n_{d};s)_{|_{s=0}}}.
\end{equation}
The final expression is given by
\begin{eqnarray}\label{mfpt_mutiple}
&& \tau(n_{1},n_{2},...,n_{d})=\frac{1}{\sqrt{k^{2}+4dku}}[-\frac{2du+k}{\sqrt{k^{2}+4dku}} \nonumber\\
&& + \frac{\Phi_{1}^{d}(2du+k+\sqrt{k^{2}+4dku})z^{\mid n_{1} \mid+\mid n_{2} \mid+...+\mid n_{d}\mid - dR }+\Phi_{2}^{d}(2du+k-\sqrt{k^{2}+4dku})z^{-\mid n_{1} \mid-\mid n_{2} \mid-...-\mid n_{d} \mid+ dR}}{\Phi_{1}^{d}(k+\sqrt{k^{2}+4dku})z^{\mid n_{1} \mid+\mid n_{2} \mid+...+\mid n_{d} \mid- dR}-\Phi_{2}^{d}(k-\sqrt{k^{2}+4dku})z^{-\mid n_{1} \mid-\mid n_{2} \mid-...-\mid n_{d} \mid+ dR }}\nonumber\\
&& + \frac{\Phi_{1}^{d}(\mid n_{1} \mid+\mid n_{2} \mid+...+\mid n_{d} \mid- dR )(2du+k+\sqrt{k^{2}+4dku})z^{\mid n_{1} \mid+\mid n_{2} \mid+...+\mid n_{d} \mid-  dR }}{\Phi_{1}^{d}(k+\sqrt{k^{2}+4dku})z^{\mid n_{1} \mid+\mid n_{2} \mid+...+\mid n_{d} \mid- dR}-\Phi_{2}^{d}(k-\sqrt{k^{2}+4dku})z^{-\mid n_{1} \mid-\mid n_{2} \mid-...-\mid n_{d} \mid+ dR }}
\nonumber\\
&& + \frac{\Phi_{2}^{d}(\mid n_{1} \mid+\mid n_{2} \mid+...+\mid n_{d} \mid- dR )(2du+k-\sqrt{k^{2}+4dku})z^{-\mid n_{1} \mid-\mid n_{2} \mid-...-\mid n_{d} \mid+ dR }}{\Phi_{1}^{d}(k+\sqrt{k^{2}+4dku})z^{\mid n_{1} \mid+\mid n_{2} \mid+...+\mid n_{d} \mid- dR}-\Phi_{2}^{d}(k-\sqrt{k^{2}+4dku})z^{-\mid n_{1} \mid-\mid n_{2} \mid-...-\mid n_{d} \mid+ dR}}\nonumber\\
&& + \frac{\Omega_{1}(2du+k+\sqrt{k^{2}+4dku})z^{\mid n_{1} \mid+\mid n_{2} \mid+...+\mid n_{d} \mid- dR}-\Omega_{2}(2du+k-\sqrt{k^{2}+4dku})z^{-\mid n_{1} \mid-\mid n_{2} \mid-...-\mid n_{d} \mid+ dR}}{\Phi_{1}^{d}(k+\sqrt{k^{2}+4dku})z^{\mid n_{1} \mid+\mid n_{2} \mid+...+\mid n_{d} \mid- dR}-\Phi_{2}^{d}(k-\sqrt{k^{2}+4dku})z^{-\mid n_{1} \mid-\mid n_{2} \mid-...-\mid n_{d} \mid+  dR}}];\nonumber\\
\end{eqnarray}
where
\begin{eqnarray}{ll}\label{Gamma}
z_{1}=\frac{(2du+k+\sqrt{k^{2}+4duk})}{2du}, &  z_{2}=\frac{(2du+k-\sqrt{k^{2}+4duk})}{2du}; &\\
\Phi_{1}=(\frac{z_{1}^{R+1}+z_{1}^{R}-2}{{z_{1}-1}}), &  \Phi_{2}=(\frac{z_{2}^{R+1}+z_{2}^{R}-2}{{z_{2}-1}}); & \\
\Omega_{1}=d\Phi_{1}^{d-1}[\frac{Rz_{1}^{R+2}-Rz_{1}^{R}-2z_{1}^{R+1}+2z_{1}}{(z_{1}-1)^2}],& \Omega_{2}=d\Phi_{2}^{d-1}[\frac{-Rz_{2}^{R+2}+Rz_{2}^{R}+2z_{2}^{R+1}-2z_{2}}{(z_{2}-1)^2}]. & 
\end{eqnarray}
The auxiliary functions $\Phi_{1}$,$\Phi_{2}$,$\Omega_{1}$ and $\Omega_{2}$ depend on the range of the source production. In the case of $R=0$ we obtain $\Phi_{1}=\Phi_{2}=1$ and $\Omega_{1}=\Omega_{2}=0$, and  Eq.(\ref{mfpt_mutiple}) reduces, as expected, to Eq. (\ref{mfpt_single}).

\newpage

\noindent Fig.1. A schematic of the discrete-state SDD model for establishment of morphogen gradients in $d$ dimensions. A specific case of $d=2$ is presented. Signaling molecule are generated at the origin (shown in red) with a rate $Q$. Particles can also diffuse along the lattice to the neighboring sites with a rate $u$, or they might be degraded with a rate $k$.

\vspace{5mm}

\noindent Fig. 2. Local accumulation times in one dimension as a function of distance from the source $r$ for discrete-state and continuum SDD models. (a) Fast degradation rates, $k=1$, $u=0.01$; (b) Comparable diffusion and degradation rates, $k=u=1$; and (c) Fast diffusion rates, $k=1$, $u=100$. The predictions for the continuum model are taken from Refs. \cite{gordon13,ellery13}. Insets show the same plots for larger length scales.
    
\vspace{5mm}

\noindent Fig. 3.  Local accumulation times in two dimensions as a function of distance from the source $r$ for discrete-state and continuum SDD models. (a) Fast degradation rates, $k=1$, $u=0.01$; (b) Comparable diffusion and degradation rates, $k=u=1$; and (c) Fast diffusion rates, $k=1$, $u=100$. The predictions for the continuum model are taken from Ref. \cite{ellery13}. Insets show the same plots for larger length scales.

\vspace{5mm}

\noindent Fig. 4. Local accumulation times in three dimensions as a function of distance from the source $r$ for discrete-state and continuum SDD models. (a) Fast degradation rates, $k=1$, $u=0.01$; (b) Comparable diffusion and degradation rates, $k=u=1$; and (c) Fast diffusion rates, $k=1$, $u=100$. The predictions for the continuum model are taken from Refs. \cite{gordon13,ellery13}. Insets show the same plots for larger length scales.

\vspace{5mm}  
    
\noindent Fig. 5. Local accumulation times at the position $r=10$ as a function of spatial dimensions. Upper curve corresponds to the  fast degradation rates, $k=1$, $u=0.01$. The middle curve is for comparable diffusion and degradation rates, $k=u=1$. The lower curve describes the fast diffusion regime, $k=1$, $u=100$. 

\vspace{5mm}

\noindent Fig. 6. Local accumulation times at the origin $r=0$ as a function of spatial dimensions. Upper curve corresponds to the  fast degradation rates, $k=1$, $u=0.01$. The middle curve is for comparable diffusion and degradation rates, $k=u=1$. The lower curve describes the fast diffusion regime, $k=1$, $u=100$. 

\vspace{5mm}

\noindent Fig. 7. The ratio of MFPT over LAT as a function of distance from the source for different dimensions  for the discrete-state  SDD models. (a) Fast degradation rates, $k=1$, $u=0.01$; (b) Comparable diffusion and degradation rates, $k=u=1$; and (c) Fast diffusion rates, $k=1$, $u=100$. 

\vspace{5mm}

\noindent Fig. 8. The ratio of variance over LAT as a function of distance from the source for different dimensions  for the discrete-state  SDD models. (a) Fast degradation rates, $k=1$, $u=0.01$; (b) Comparable diffusion and degradation rates, $k=u=1$; and (c) Fast diffusion rates, $k=1$, $u=100$. 

\vspace{5mm}

\noindent Fig. 9. Local accumulation times as a function of the size of the production region at the source surface, $r=R\sqrt{d}$. The size of the source region is expressed in units of $R\sqrt{d}$. (a) Fast degradation rates, $k=1$, $u=0.01$; (b) Comparable diffusion and degradation rates, $k=u=1$; and (c) Fast diffusion rates, $k=1$, $u=100$. 

\newpage

\begin{figure}[h]
\begin{center}
\unitlength 1in
  \resizebox{3.375in}{2in}{\includegraphics{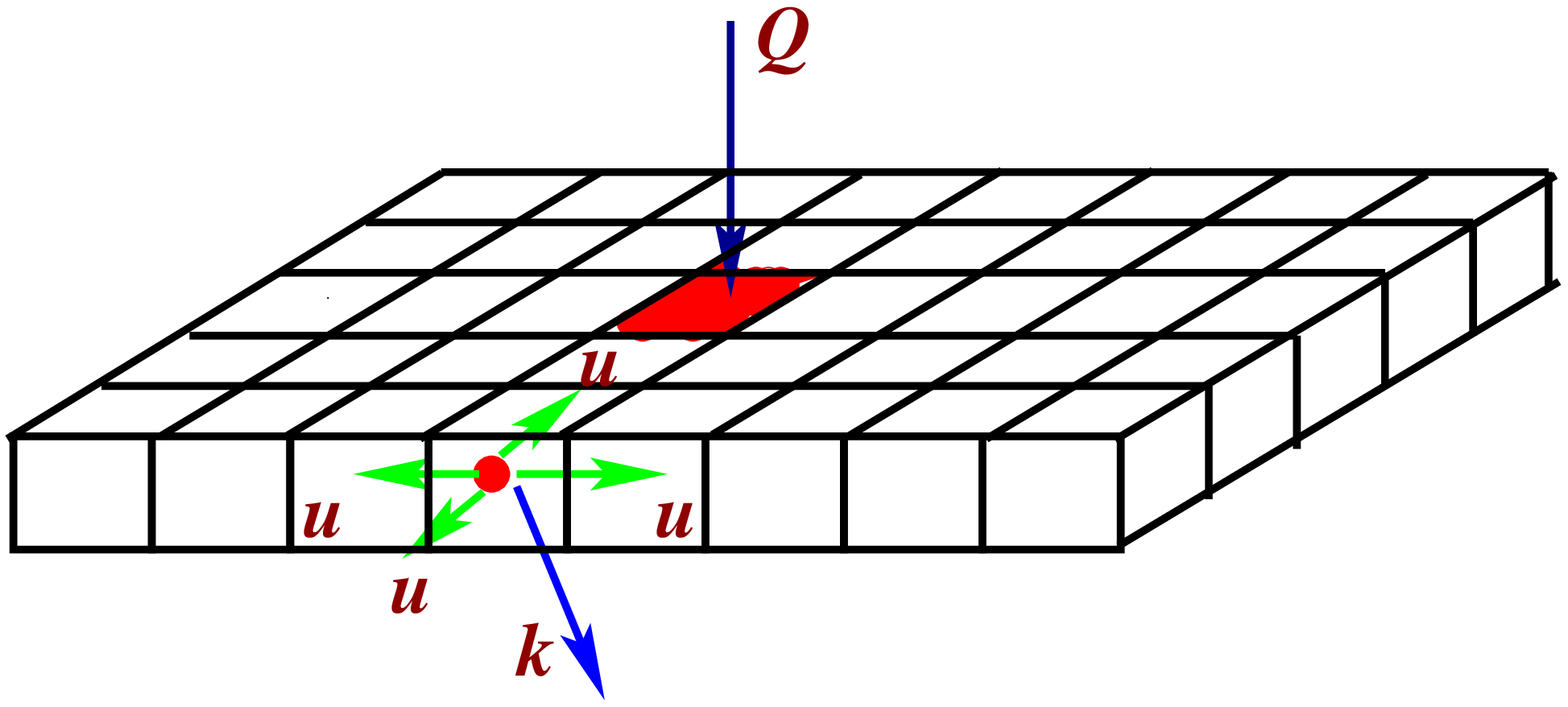}}
\vskip 1in
\begin{Large} Figure 1. Teimouri and Kolomeisky \end{Large}
\end{center}
\end{figure}

\newpage

\begin{figure}[h]
\begin{center}
\unitlength 1in
  \resizebox{3.375in}{3.375in}{\includegraphics{fig2.eps}}
\vskip 1in
\begin{Large} Figure 2. Teimouri and Kolomeisky \end{Large}
\end{center}
\end{figure}
  
\newpage

\begin{figure}[h]
\begin{center}
\unitlength 1in
  \resizebox{3.375in}{3.375in}{\includegraphics{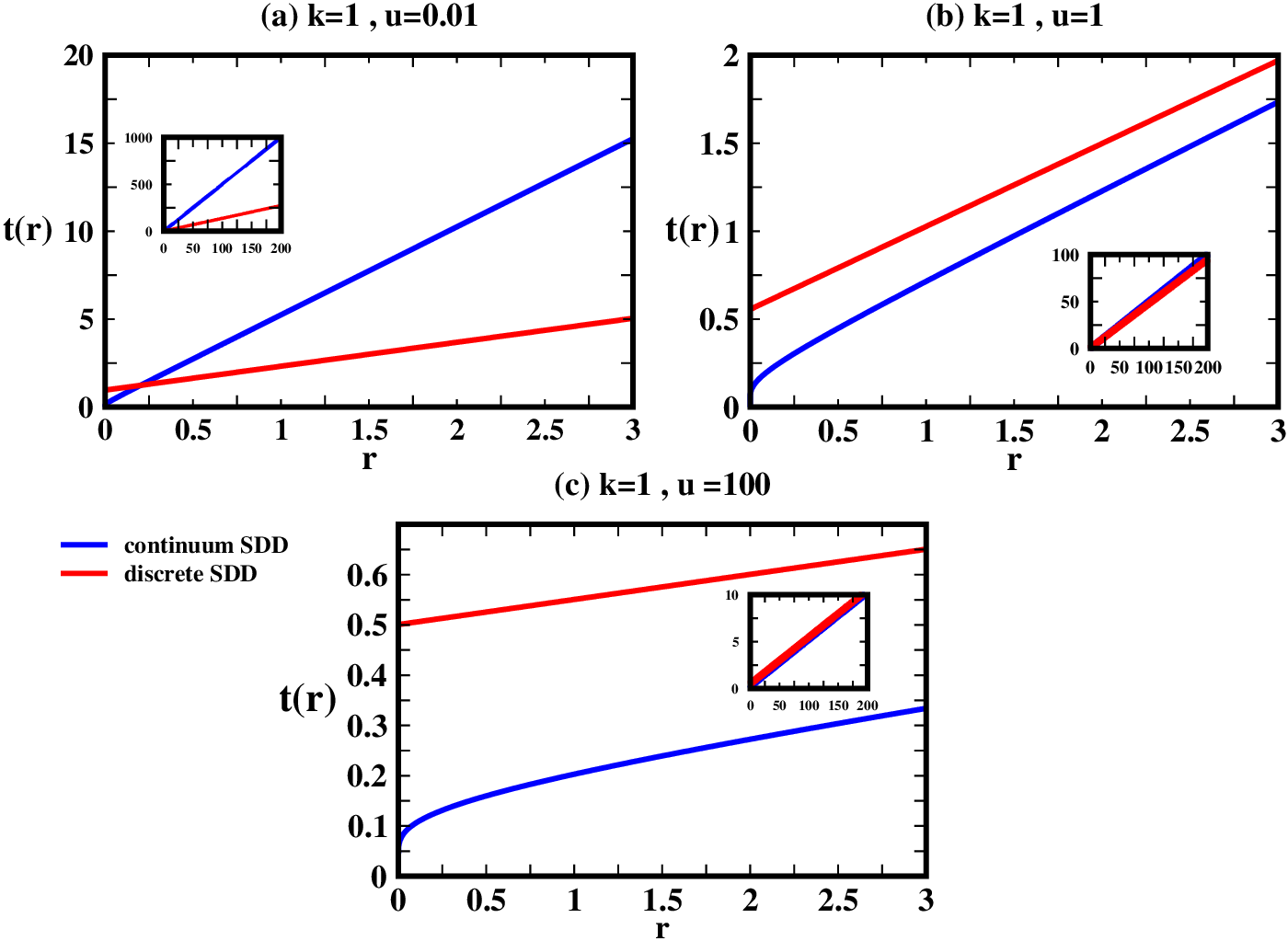}}
\vskip 1in
\begin{Large} Figure 3. Teimouri and Kolomeisky \end{Large}
\end{center}
\end{figure}

\newpage

\begin{figure}[h]
\begin{center}
\unitlength 1in
  \resizebox{3.375in}{3.375in}{\includegraphics{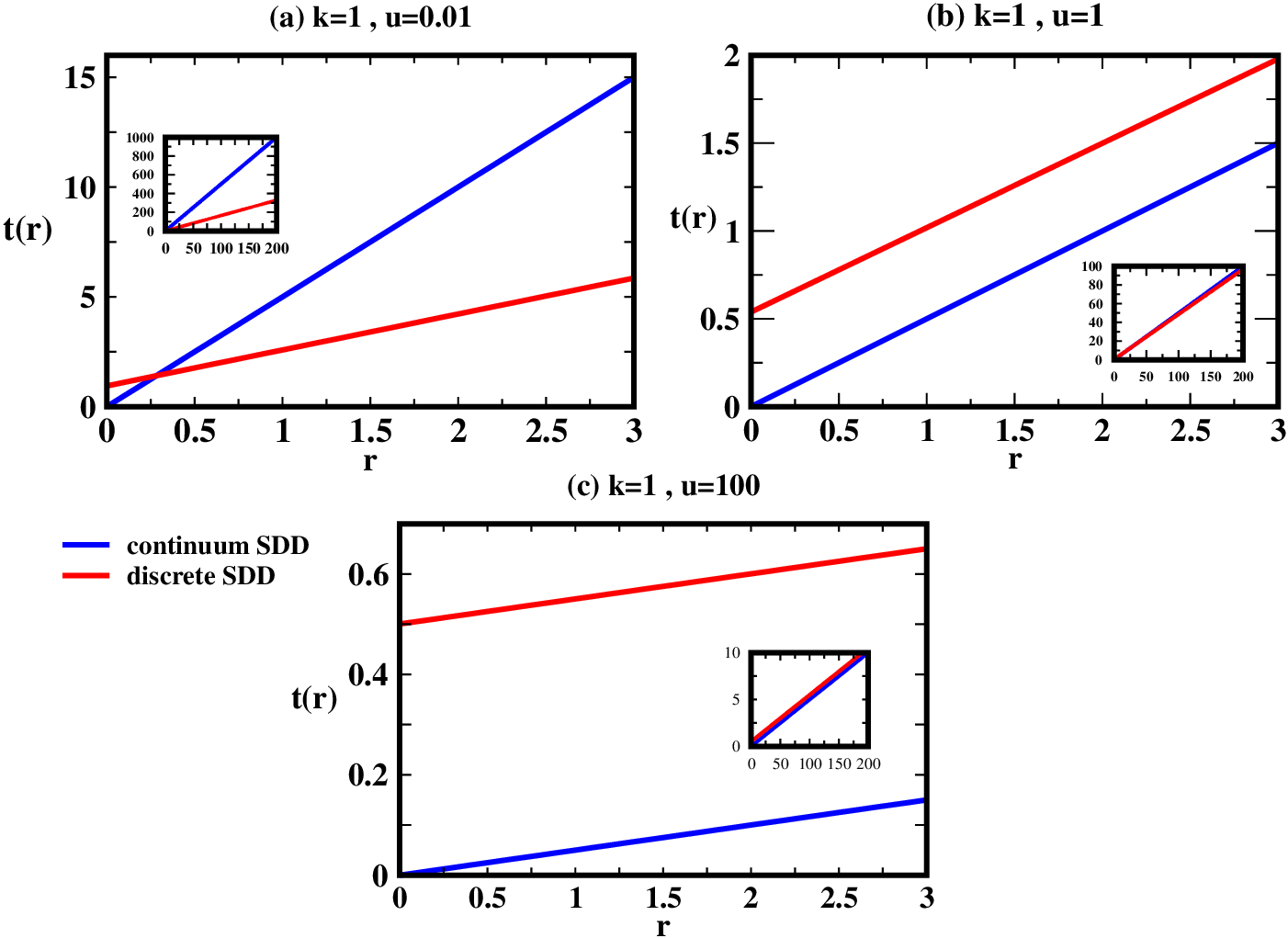}}
\vskip 1in
\begin{Large} Figure 4. Teimouri and Kolomeisky \end{Large}
\end{center}
\end{figure}

\newpage

\begin{figure}[h]
\begin{center}
\unitlength 1in
  \resizebox{3.375in}{3.375in}{\includegraphics{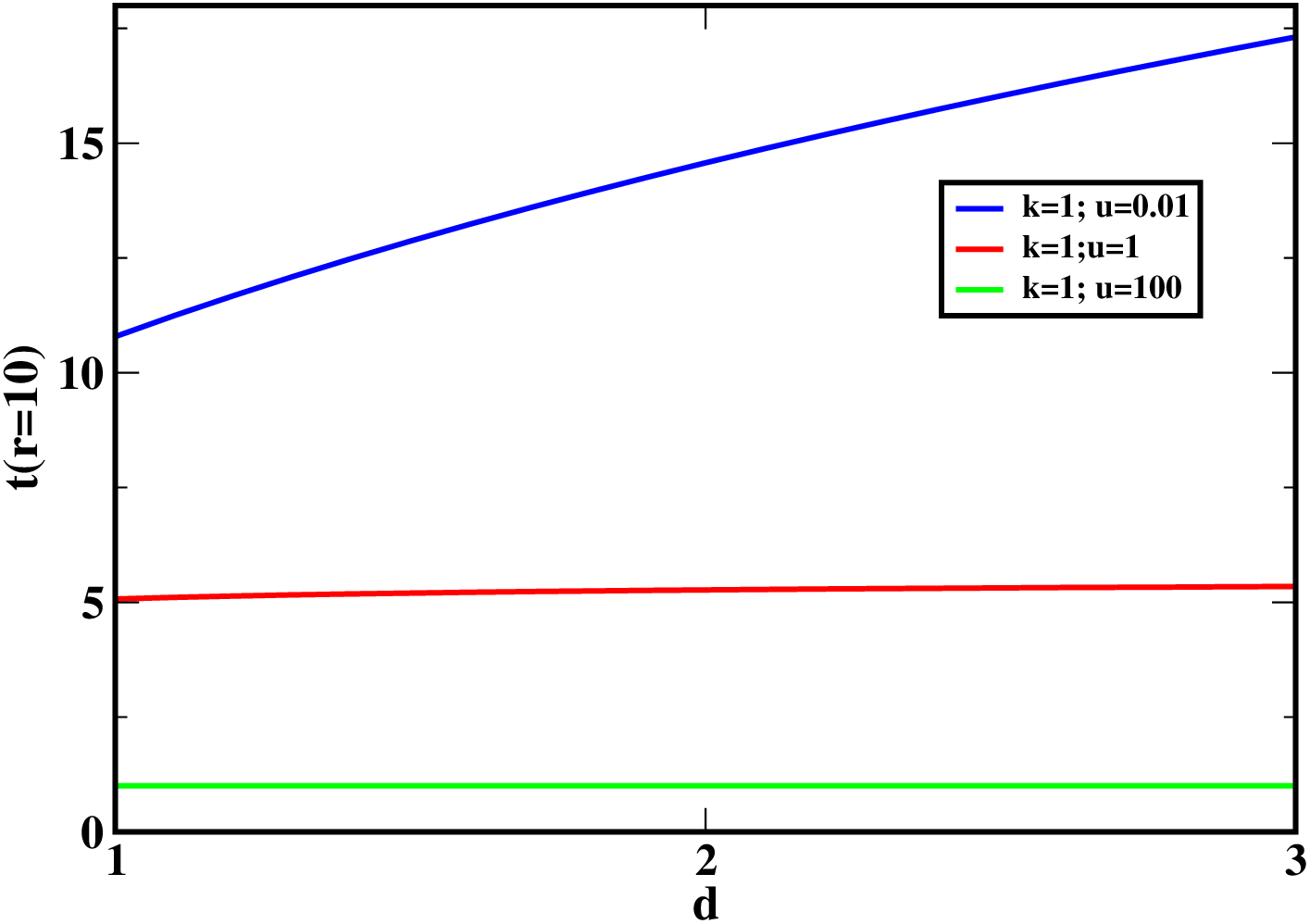}}
\vskip 1in
\begin{Large} Figure 5. Teimouri and Kolomeisky \end{Large}
\end{center}
\end{figure}

\newpage

\begin{figure}[h]
\begin{center}
\unitlength 1in
  \resizebox{3.375in}{3.375in}{\includegraphics{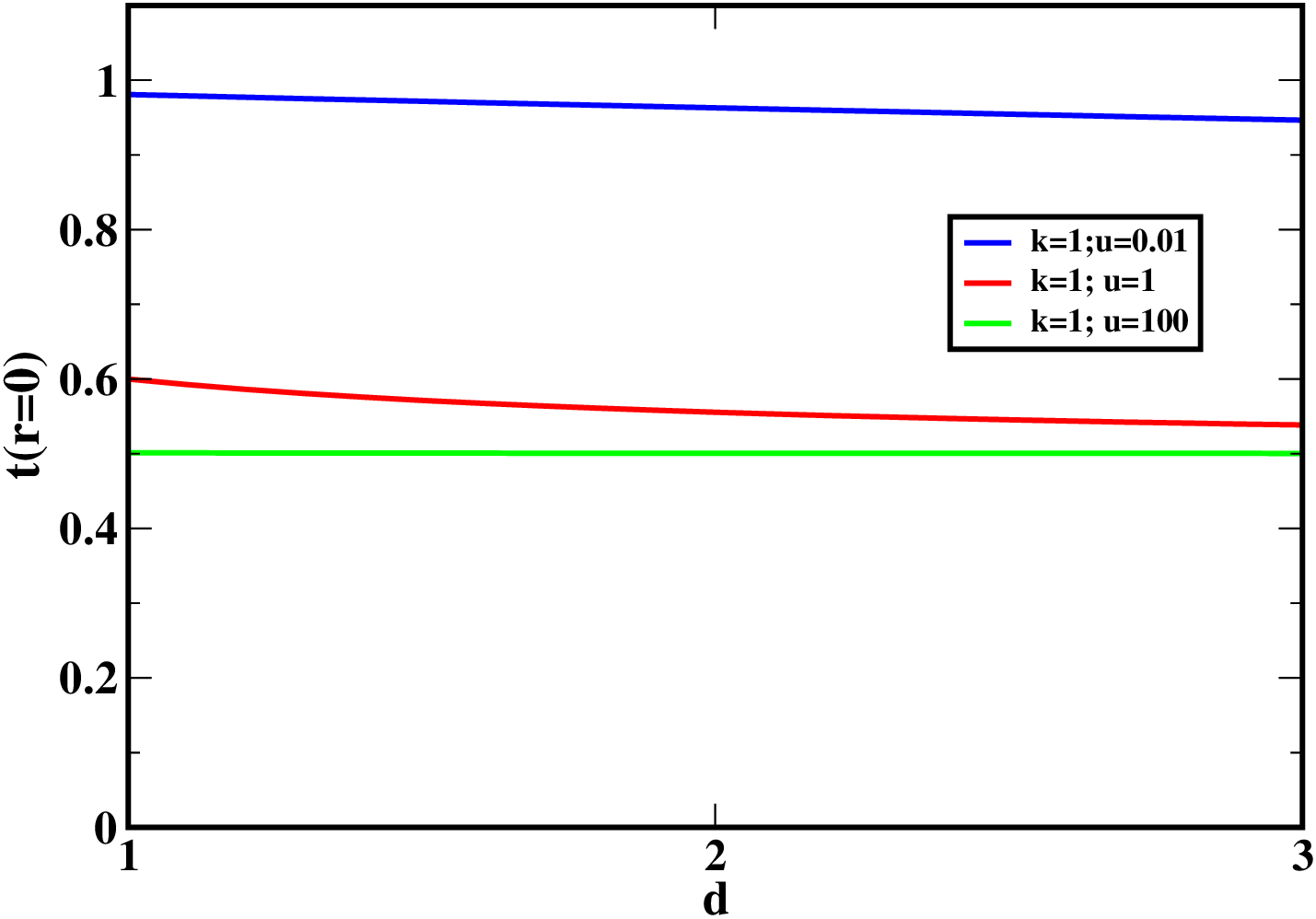}}
\vskip 1in
\begin{Large} Figure 6. Teimouri and Kolomeisky \end{Large}
\end{center}
\end{figure}

\newpage

\begin{figure}[h]
\begin{center}
\unitlength 1in
  \resizebox{3.375in}{3.375in}{\includegraphics{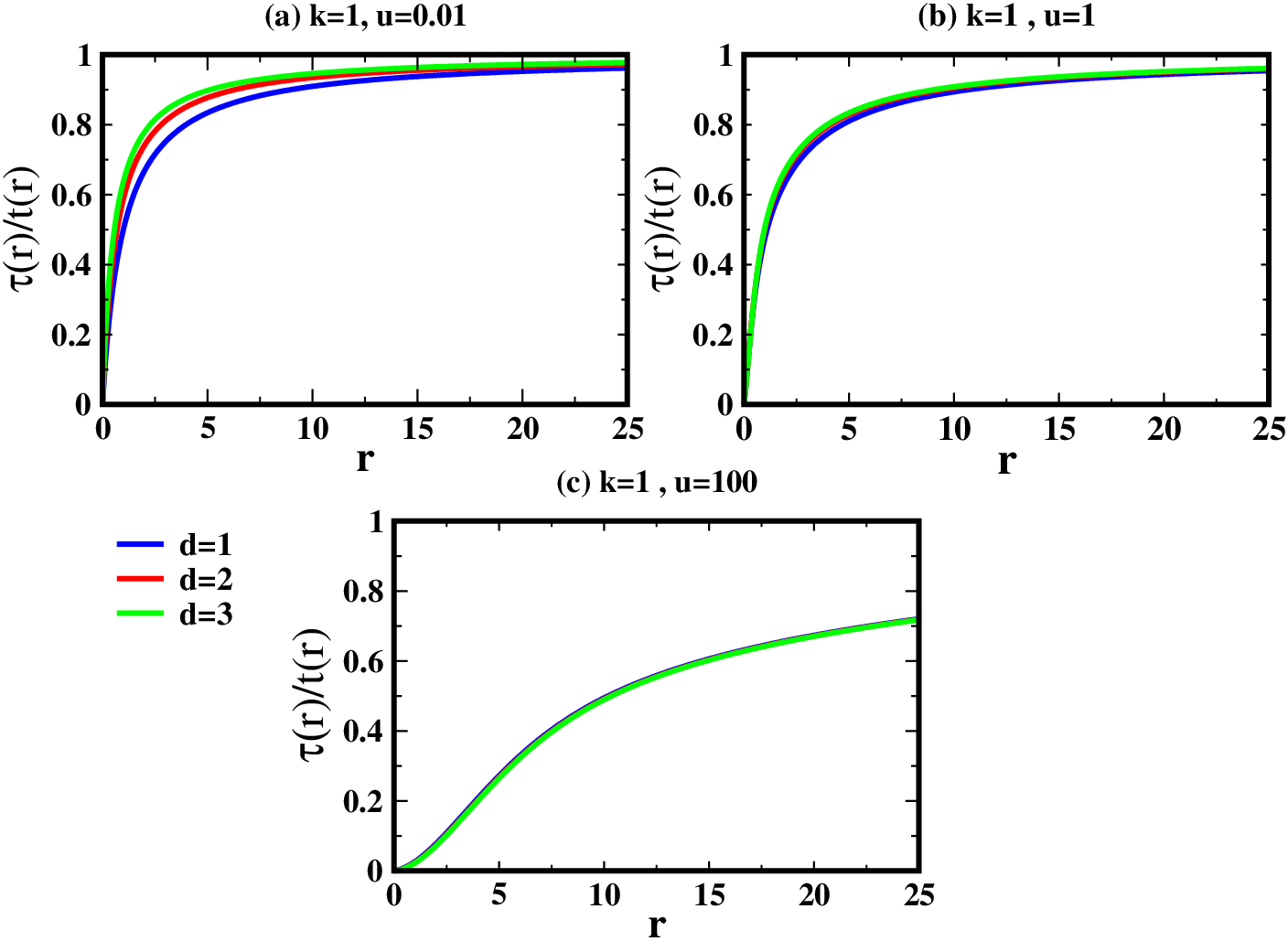}}
\vskip 1in
\begin{Large} Figure 7. Teimouri and Kolomeisky \end{Large}
\end{center}
\end{figure}

\newpage

\begin{figure}[h]
\begin{center}
\unitlength 1in
  \resizebox{3.375in}{3.375in}{\includegraphics{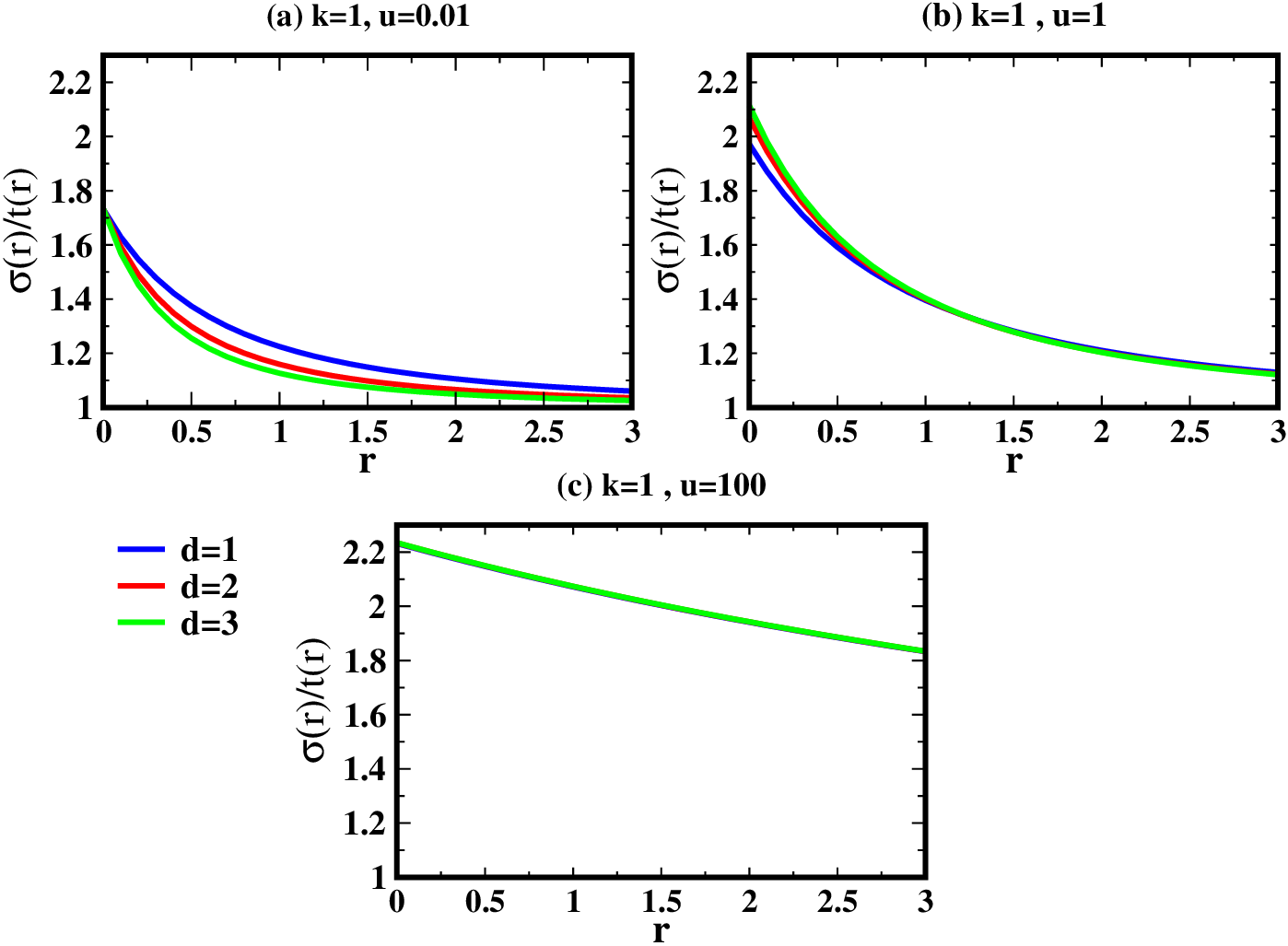}}
\vskip 1in
\begin{Large} Figure 8. Teimouri and Kolomeisky \end{Large}
\end{center}
\end{figure}

\newpage

\begin{figure}[h]
\begin{center}
\unitlength 1in
  \resizebox{3.375in}{3.375in}{\includegraphics{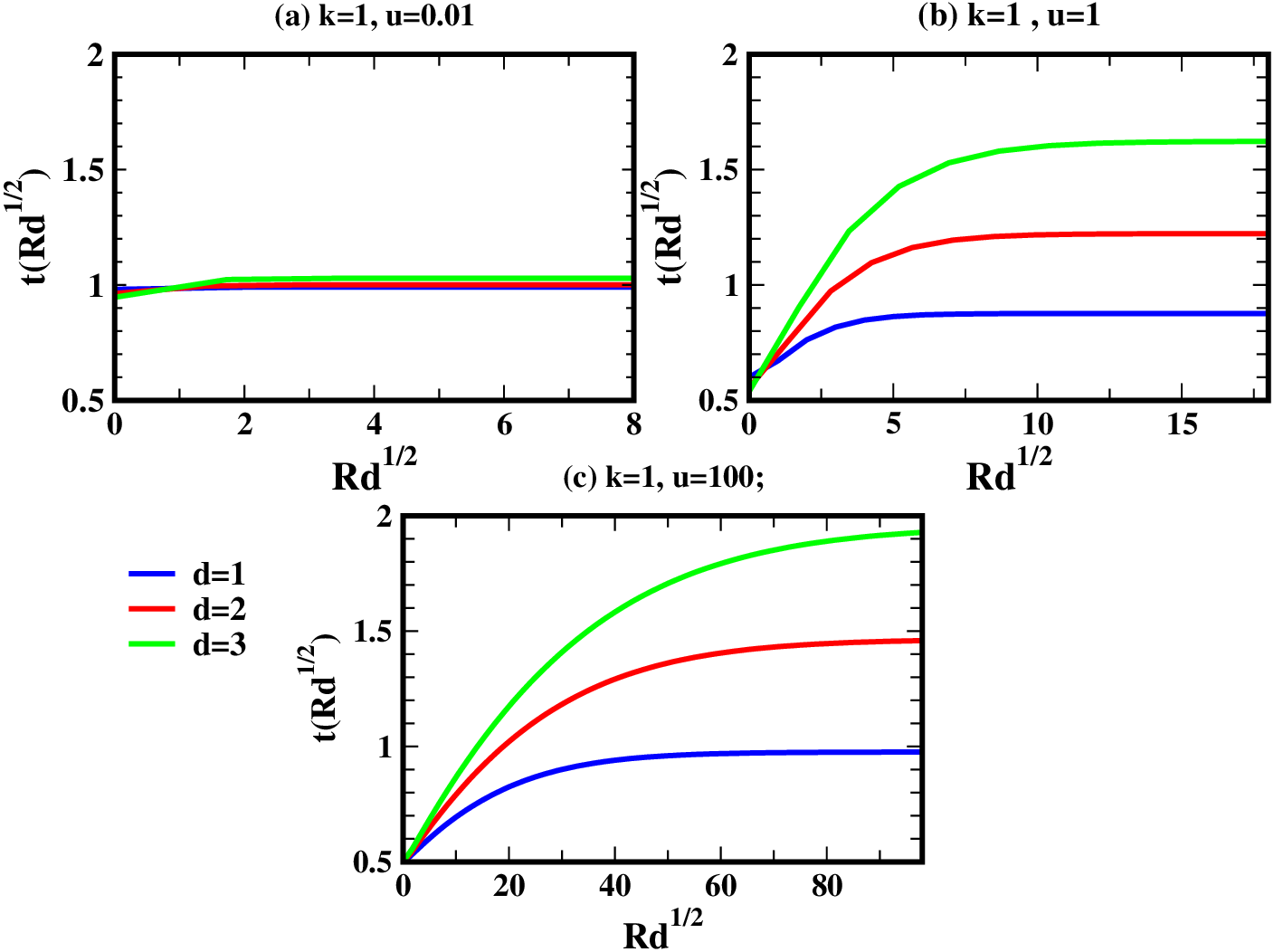}}
\vskip 1in
\begin{Large} Figure 9. Teimouri and Kolomeisky \end{Large}
\end{center}
\end{figure}


\begin{thebibliography}{99}



\bibitem{book}  A. Martinez-Arias and  A. Stewart,  {\it Molecular Principles of Animal Development}  (Oxford University Press, New York, 2002).

\bibitem{lodish_book} H. Lodish, A. Berk, C.A. Kaiser, M. Krieger, M.P. Scott, A. Bretscher, H. Ploegh, and  P. Matsudaira, {\it Molecular Cell Biology} 6-th ed., (W.H. Freeman, New York, 2007).

\bibitem{wolpert69} L. Wolpert, J. Theor. Biol. {\bf 25}, 1 (1969).

\bibitem{tabata04} T. Tabata and Y. Takei,  Development {\bf 131}, 703 (2004).

\bibitem{crick70}  F.H. Crick,  Nature {\bf 225}, 420 (1970).

\bibitem{turing}  A.M. Turing, Phil. Trans. Roy. Soc. Lond. {\bf 237}, 37 (1952).

\bibitem{porcher10} A. Porcher and N. Dostatni,  Curr. Biol.  {\bf 20}, R249 (2010).

\bibitem{gregor07} T. Gregor, E.F. Wieschaus, A.P. McGregor, W. Bialek and D.W. Tank,  Cell {\bf 130}, 141 (2007).

\bibitem{kicheva07} A. Kicheva, P. Pantazis, T. Bollenbach, Y. Kalaidzidis, T. Bittig, F. J\"{u}licher and  M. Gonzales-Gaitan,  Science {\bf 315}, 521 (2007).

\bibitem{yu09} S.R. Yu, M. Burkhardt, M. Nowak, J. Ries, Z. Petrasek, S. Scholpp, P. Schwille and M. Brand,  Nature {\bf 461}, 533 (2009).

\bibitem{kerszberg98} M. Kerszberg and L. Wolpert, J. Theor. Biol. {\bf 191}, 103 (1998).

\bibitem{entchev00}  E.V. Entchev, A. Schwabedissen and M. Gonzales-Gaitan,  Cell {\bf 103}, 981 (2000).

\bibitem{muller12} P. M\"{u}ller, K. W. Rogers, B. M. Jordan, J. S. Lee, D. Robson, S. Ramanathan, A. F. Schier,  Science {\bf 336}, 721 (2012).

\bibitem{drocco11} J.A. Drocco, O. Grimm, D.W. Tank and E. Wieschaus, Biophys. J. {\bf 101}, 1807 (2011).

\bibitem{little11}  S. C. Little,  G. Tkacik,  T. B. Kneeland,  E. F. Wieschaus,  T. Gregor, PLoS Biol. {\bf 9} e1000596 (2011) 

\bibitem{spirov09} A. Spirov, K. Fahmy, M. Schneider, E. Frei, M. Noll and S. Baumgartner, Development 
 {\bf 136} 605-614 (2009)

\bibitem{zhou12} S. Zhou, W. C. Lo, J. L. Suhalim, M. A. Digman, E. Grattom, Q. Nie and A. D. Lander, Current Biology {\bf 22} 668-675 (2012)


\bibitem{kornberg12} T.B. Kornberg, Biophys. J. {\bf 103}, 2252 (2012).

\bibitem{drocco12} J.A. Drocco, E. F. Wieschaus and D. W. Tank, Phys. Biol. {\bf 9}, 055004 (2012).

\bibitem{berezhkovskii10} A.M. Berezhkovskii, C. Sample, and S.Y. Shvartsman,  Biophys. J. {\bf 99}, L59 (2010).

\bibitem{berezhkovskii11a}  A.M. Berezhkovskii,  J. Chem. Phys. {\bf 135}, 07412 (2011).

\bibitem{berezhkovskii11b} A.M. Berezhkovskii and S.Y. Shvartsman, J. Chem. Phys. {\bf 135}, 154115 (2011).

\bibitem{berezhkovskii11c} A.M. Berezhkovskii, C. Sample and S.Y. Shvartsman, Phys. Rev. E {\bf 83}, 051906 (2011). 

\bibitem{berezhkovskii13} A. M. Berezhkovskii and   A.M. Shvartsman J. Chem. Phys. {\bf 138}, 244105(2013).

\bibitem{mogilner11}  A. Mogilner and D. Odde, Trends Cell Biol.  {\bf 21}, 692-700 (2011).

\bibitem{muller13}  P. M\"{u}ller, K. W. Rogers, S. R. Yu, M. Brand, A. F. Schier, Development  {\bf 140}, 1621 (2013).

\bibitem{gordon13}  P. V. Gordon, C. B. Muratov, and S. Y. Shvartsman , J. Chem. Phys. {\bf 138}, 104121 (2013).

\bibitem{ellery13}  A. J. Ellery, M. J. Simpson, and S. W. McCue , J. Chem. Phys. {\bf 139}, 017101 (2013).

\bibitem{kolomeisky11}  A.B. Kolomeisky, J. Phys. Chem. Lett. {\bf 2}, 1502 (2011).

\bibitem{redner}  S. Redner,{\it A Guide to First-Passage Processes} (Cambridge University Press, New York, 2001).













\end{thebibliography}
\end{document}